\documentclass[preprints,review,accept,oneauthor,pdftex]{Definitions/mdpi} 

\firstpage{1} 
\pubvolume{1}
\issuenum{1}
\articlenumber{0}
\pubyear{2021}
\copyrightyear{2020}
\datereceived{} 
\dateaccepted{} 
\datepublished{} 
\hreflink{https://doi.org/} 

\newcommand{\collend}{\end{paracol}}

\Title{TeV Instrumentation: current and future}

\TitleCitation{TeV Instrumentation: current and future}

\Author{Julian Sitarek$^{1}$\orcidA{}}

\AuthorNames{Julian Sitarek}

\AuthorCitation{Sitarek, J.}

\address{%
$^{1}$ \quad  Institute for Cosmic Ray Research (ICRR), The University of Tokyo, Kashiwa, 277-8582 Chiba
}

\abstract{
  During the last 20 years, TeV astronomy turned from a fledgling field, with only a handful of sources into a fully-developed astronomy discipline, broadening our knowledge on a variety of types of TeV gamma-ray sources.
  This progress has been mainly achieved due to currently operating instruments: Imaging Atmospheric Cherenkov Telescopes, Surface Array and Water Cherenkov  detectors.
  Moreover, we are at the brink of a next generation of instruments, with a considerable leap of performance parameters.
  This review summarises the current status of the TeV astronomy instrumentation, mainly focusing on the comparison of the different types of instruments and analysis challenges, as well as provides an outlook into the future installations.
  The capabilities and limitations of different techniques of observations of TeV gamma rays are discussed, as well as synergies to other bands and messengers. 
}


\begin{document}


\section{Introduction}

Out of the different messengers (see Chapter 7 of this Special Issue) VHE (very-high-energy, $\gtrsim100$\,GeV) gamma rays so far have been the most successful in investigating the most extreme, high-energy  processes in extragalactic sources.
Cosmic rays, except of those exceeding EeV energies, are isotropised by galactic and extragalactic magnetic fields and hence cannot be pinpointed to individual sources.
Neutrino astronomy, while being able to unambiguously point out occurrence of hadronic processes in the source, is severely marred by very small statistics of detectable events.
Gamma-ray astronomy solves both of those problems, as the high-energy photons can be easily detected by their interactions with the matter, and cosmic gamma-ray sources produce emission associated with particular direction in the sky.

Due to the interaction of gamma rays with the Earth's atmosphere, they cannot be observed directly by ground-based instruments.
Instead, balloon-based (see e.g. GRAINE, \citealp{Takahashi:2021qj}), or particularly successful space-borne (see e.g. Large Area Telescope on board of the \textit{Fermi} satellite, \citealp{2009ApJ...697.1071A}) instruments are effective in monitoring the sky at the GeV energies.
Nevertheless, at VHE range such a technique is no longer efficient.
As the balloon and satellite experiments have the collection area comparable to their physical size (of the order of m$^2$), even for bright sources this allows detection of only of the order of a single event per day above 100\,GeV.
Such low event rates prevents investigations of short term phenomenons in this energy range with direct detection of gamma rays.

\begin{figure}[t]
    \centering
    \includegraphics[width=0.6\textwidth]{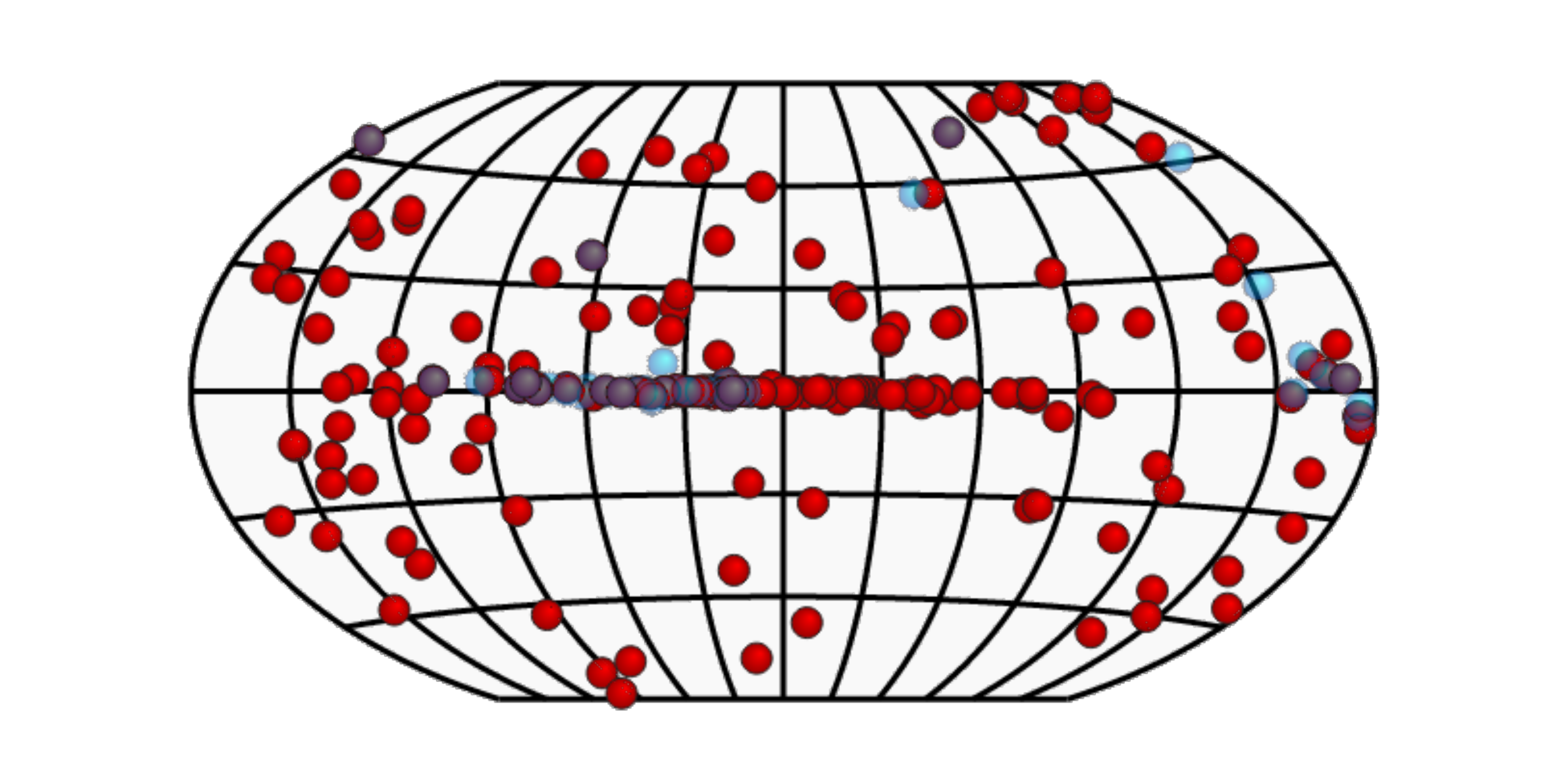}
    \caption{Overlay image in Galactic coordinates of the sources seen by the currently operating major IACT experiments (H.E.S.S., MAGIC, VERITAS, in red) and by currently operating WCD and SA experiments (HAWC, LHAASO, in blue). Sources detected by both techniques show up in violet. 
    Data obtained using TeVCAT (\url{http://tevcat2.uchicago.edu}) service. }
    \label{fig:sources}
\end{figure}

Thankfully, the absorption of gamma rays in the atmosphere can be turned into an indirect method of detecting radiation in the VHE range.
A primary gamma ray converts into a $e^+e^-$ pair in the radiation field of an atmospheric nuclei.
Those in turn produce further gamma rays in the Bremsstrahlung process.
The combination of both processes results in an electromagnetic cascade propagating through the atmosphere.
The cascade, often referred to as an (extensive) atmospheric shower,  initiated by a TeV gamma ray is composed of thousands of highly relativistic particles. 
Such events can be observed directly by detecting the particles surviving till the ground level with a Surface Arrays (SA) or Water Cherenkov Detectors (WCD).
Alternatively, the events can be studied by observations of Cherenkov light emitted by particles during the shower development (Imaging Atmospheric Cherenkov Telescope technique, IACT). 
Instruments exploiting both of those techniques allowed to expand the catalogue of VHE sources up to $\sim250$ objects (see Fig.~\ref{fig:sources}). 
While IACT technique was used to detect both Galactic and extragalactic sources, so far most of the SA/WCD sources have been claimed in the Galactic plane.  
In this review both techniques are discussed, their advantages/disadvantages and complementarity. 
The current and future instruments are covered as well as connection to other bands and messengers.

\section{Instrumentation}

\begin{figure}[t]
    \centering
    \includegraphics[width=0.6\textwidth, trim=0 150 0 50, clip]{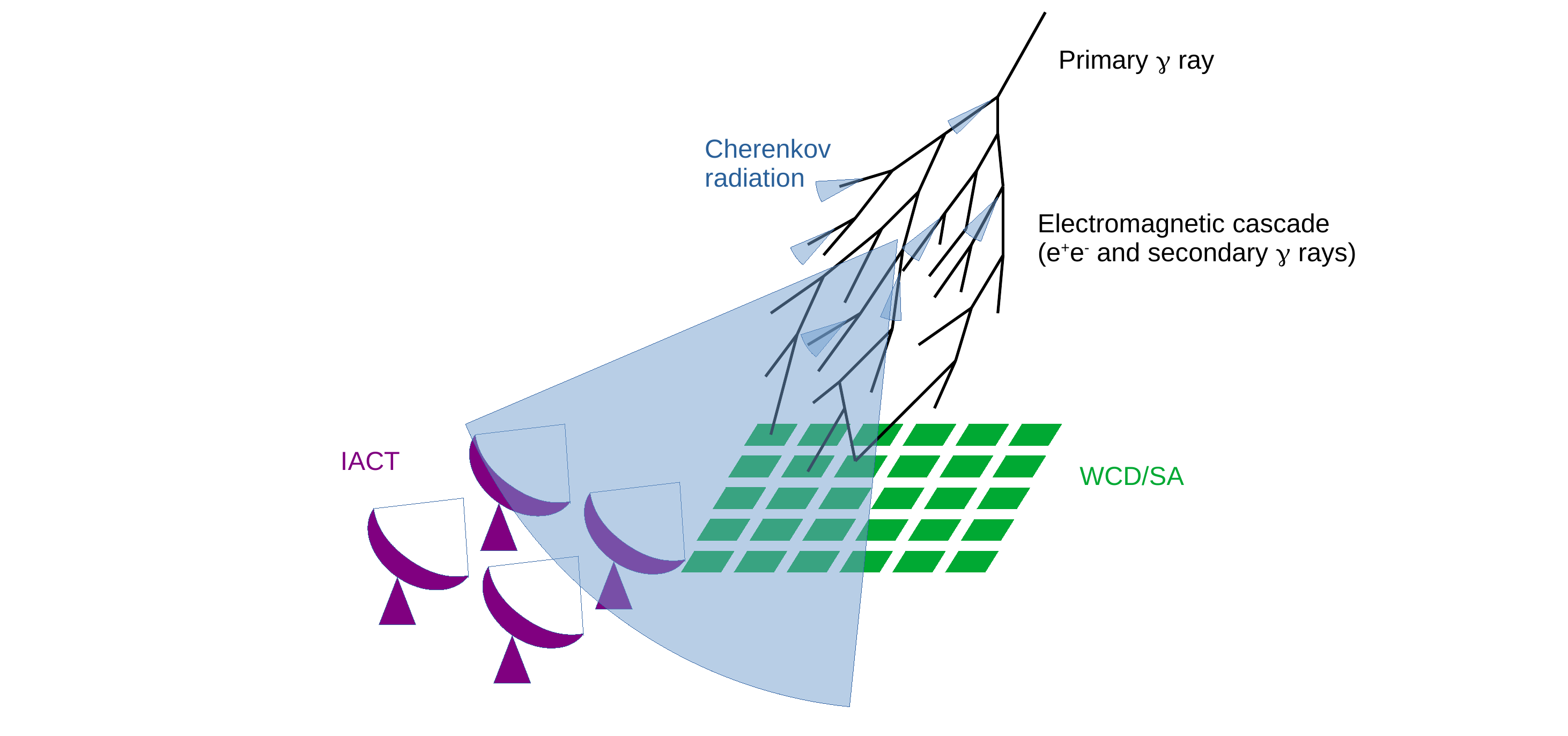}
    \caption{Principle of TeV gamma ray detection. Primary gamma ray interacts with the atmospheric nuclei creating a cascade (atmospheric shower) of secondary particles (black lines). Charged particles with their speed exceeding the speed of light in the atmosphere induce emission of the Cherenkov light flashes (blue cones). The radiation can be observed by IACTs (violet).
    Those of the secondary particles that reach the ground level can be also directly detected by WCD or SA experiments (green). }
    \label{fig:shower}
\end{figure}

Both techniques share the same principle of reconstructing the primary gamma ray events from the properties of their extensive air shower (see Fig.~\ref{fig:shower}). 
They also struggle with the same issue of dominating background due to similar showers produced by isotropic cosmic ray particles. 
Nevertheless, usage of either the Cherenkov light for tracking the development of the shower, or the remaining shower particles reaching the ground causes a number of differences between the two approaches, affecting strongly the achieved performance. 

\subsection{IACTs technique}
A shower produced by a TeV primary gamma ray contains thousands of ultra-relativistic particles. 
Charged particles moving faster than the light in the atmosphere will induce production of dim and short (of the order of nanoseconds) flashes of Cherenkov light. 
As the TeV showers reach their maximum at the height of $\sim10$\,km a.s.l., and the Cherenkov radiation is emitted at an angle of $\sim1^\circ$, the radiation is spread over a region with a radius of $\sim 120$\,m, the so-called light pool.  
A telescope located in such a light pool, can detect the Cherenkov emission. 
Hence, despite the physical sizes of the current generation IACT telescopes are few hundred of m$^2$, they can achieve effective collection area for TeV gamma rays of the order of $10^5$\,m$^2$. 

Two main components of the IACT are the mirror dish (including a drive system to reposition it) that collects the Cherenkov photons and the camera (as well as the trigger and data acquisition systems) that allows conversion of Cherenkov light into an image of the shower. 
The size of the mirror dish determines the energy threshold of an IACT.
The smaller ($\lesssim15$\,m diameter) telescopes employ the Davies-Cotton shape of the mirror dish, that minimise the optical aberrations. 
For larger telescopes ($\gtrsim15$\,m diameter) the parabolic shape is used, since it minimises the time spread of isochronous radiation reaching the telescope.
Due to large sizes of the telescope dishes, tessellated mirrors are used. 
Unless the telescope structure is stiff enough \citep{2015ICRC...34.1017G}, the changing gravitational loads as the telescope moves requires constant correction of the individual direction (see \citealp{2016APh....82....1A} and references therein) to maintain its accurate  pointing.

One of the key features that added to the success of IACTs is their ``Imaging'' capabilities of their cameras. 
The light collected by the telescope is focused on optical plane composed of the order of 1000 photosensors (most commonly photomultipliers, PMT, however Silicon Photomultipliers, SiPM, are also being considered or already used, \citealp{2012NIMPA.695...96K}).
The camera allows one to construct the two-dimensional angular distribution of the observed light, the image of the shower.
Such images can be parametrised as ellipses, which facilitates the reconstruction of the properties of the primary particle \citep{1985ICRC....3..445H}. 
Thanks to the changing atmospheric refractive index with the height a.s.l., the long axis of the ellipse tracks the longitudinal evolution of the shower.
On the other hand, the short axis is related to the latitudinal distribution.
The angular image from a single telescope is however not sufficient to fully disentangle between showers that reach their maximum at the same line of sight, but at different heights above the ground (or alternatively at different impact distances). 
Combination of images of the same event seen by multiple nearby telescopes provides a natural way to reconstruct the 3-dimensional geometry of the event (see e.g. \citealp{1999APh....12..135H}).
Stereoscopic trigger is also an efficient way of removing a background of single muon events both at hardware level (see e.g. \citealp{2006APh....25..391H}) and during the analysis (see e.g. \citealp{2012APh....35..435A}).
The three major, currently operating, stereoscopic IACT systems are H.E.S.S.\citep{2020APh...11802425A}, MAGIC \citep{2016APh....72...61A} and VERITAS \citep{2006APh....25..391H}.

\subsection{SA and WCD technique}
In contrast to IACTs, surface detectors exploit directly the particles that reach the ground level.
One way of detecting those particles is the surface arrays 
(SA\footnote{Various names are used for those types of detectors, including: surface arrays, surface detectors, air shower arrays.}) of scintillation counters (see e.g. \citealp{1996NIMPA.383..431K,1999ApJ...525L..93A}), that track passing of charged particles. 
A small layer of dense material (typically lead) placed above such a counter provides means of conversion of secondary gamma rays into $e^+e^-$ pair (as well as slightly raises the detection energy threshold for the secondary charged particles).
The net effect of such an absorption layer is improvement of the energy and angular resolutions and decrease of the energy threshold for primary particles. 

While the SA are usually finely spaced to maximise the collection area at highest energy it is also possible to construct dense, or full-coverage arrays. 
The experiment ARGO-YBJ used Resistive Plate Chambers to cover dense carpet of $72\,\mathrm{m} \times 76\,\mathrm{m}$ with 92\% coverage and an additional guard ring in the total area of $110\,\mathrm{m} \times 100\,\mathrm{m}$ \cite{2013ApJ...779...27B}.

Alternatively, WCD uses water tanks to absorb and detect secondary particles produced in the shower.
Secondary charged particle moving faster than about $0.77\,c$ will induce production of Cherenkov radiation inside such a tank, that can be gathered by large size (typically the photocatode with a diameter of over 20\,cm) PMTs. 
While SA stations basically detect the number of charged particles (rather than energy deposit), it is more complicated in the case of WCD. 
Electrons/positrons reaching a WCD detector encounter multiple radiation lengths of water that efficiently absorb them, facilitating a calorimetric measurement of this component. 
However, muons pass through the detector producing a signal that is not proportional to their energy, but rather the detector crossing length. 

The previous generation WCD such as MILAGRO \citep{1996SSRv...75..199Y} was using a single pond of water and a grid of PMTs. 
In such a setup the scattered Cherenkov radiation could travel to neighbouring PMTs blurring the response of the detector. 
HAWC \citep{2017ApJ...843...40A} and LHAASO-WCDA \citep{2021PhRvD.104f2007A} use respectively separate tanks or curtains to counteract such an effect.
Sufficiently high above the energy threshold, the collection area of the gamma rays is closely related with physical size of the array.
On the other hand, the energy threshold of the instrument is determined by the height a.s.l. of the WCD and its particle collection efficiency (which depends also on the fill factor of the array, i.e. how big fraction of the total array area is covered by the detectors). 
In order to limit the random coincidences rate, a trigger can exploit time correlation of signals of different parts of detectors depending on their relative distance (see \cite{2004ITNS...51.1835A} for ARGO-YBJ).

The content of the shower particles depends on the primary particle type.
While gamma-ray initiated showers produce nearly exclusively gamma-ray and $e^\pm$ particles, in the case of hadron-induced shower there is an additional component from muons, that are much more penetrative. 
Therefore inclusion of additional particle counters located under a natural, deep layer of absorbing material (typically earth or water) provides a measurement of the muon content of the shower that can be used in selection of gamma-ray showers  \cite{2007Ap&SS.309..435A}.

The sampling of local (and time-dependent) secondary particle distribution on the ground by WCD or SA detectors is in a sense equivalent to the imaging capabilities of IACTs.
Both methods construct an image of the shower (in the ground coordinates for SA/WCD tracking the lateral distribution of particles and in angular coordinates for IACT, tracking also the longitudinal profile of the shower) and based on it allow us to determine the structure of the atmospheric shower, and thus reconstruct the energy, arrival direction and also reduce the background from cosmic-ray-induced showers.

\subsection{Comparison of performance and synergies}

\begin{figure}[t]
    \centering
    \includegraphics[width=0.35\textwidth]{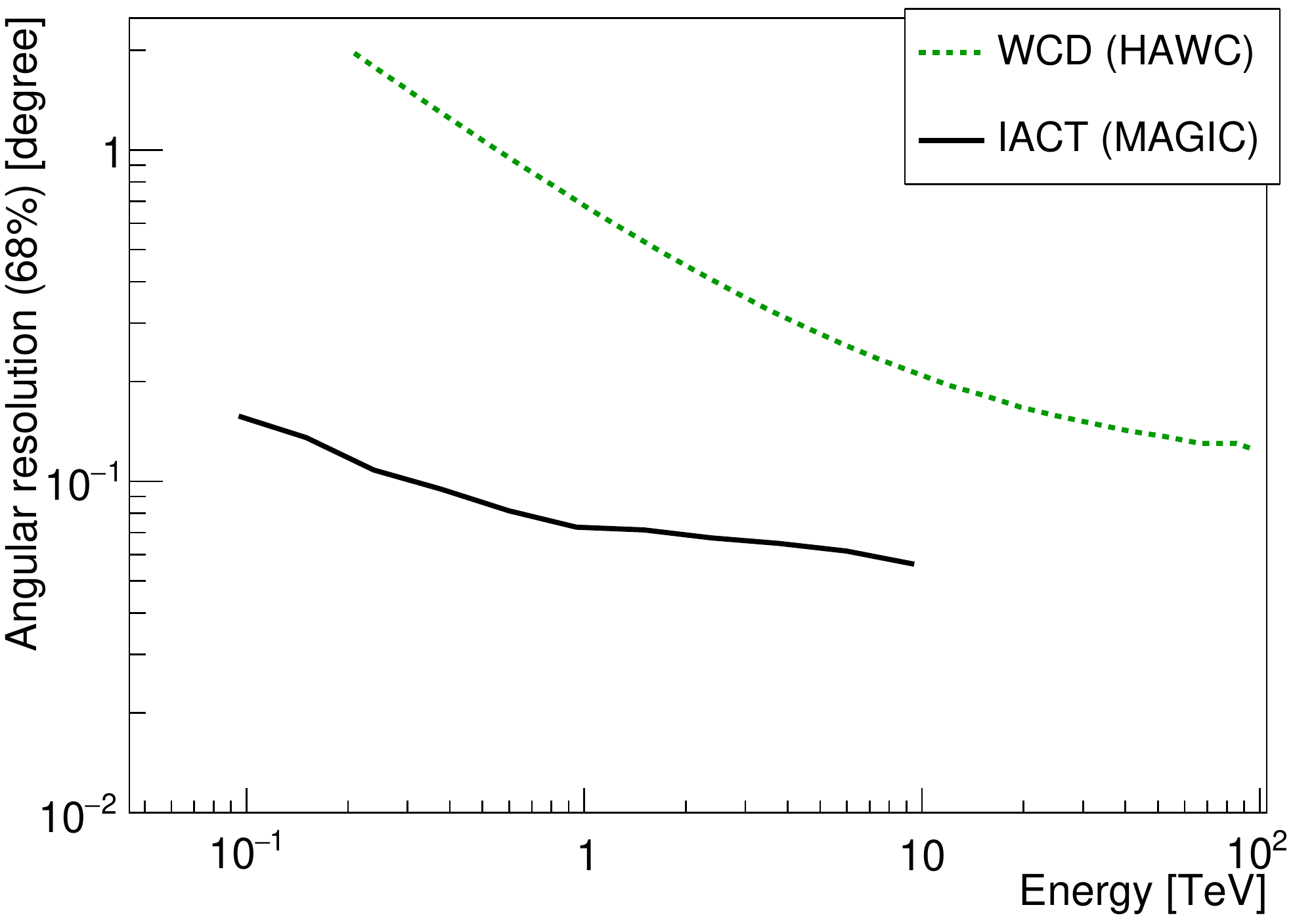}
    \includegraphics[width=0.35\textwidth]{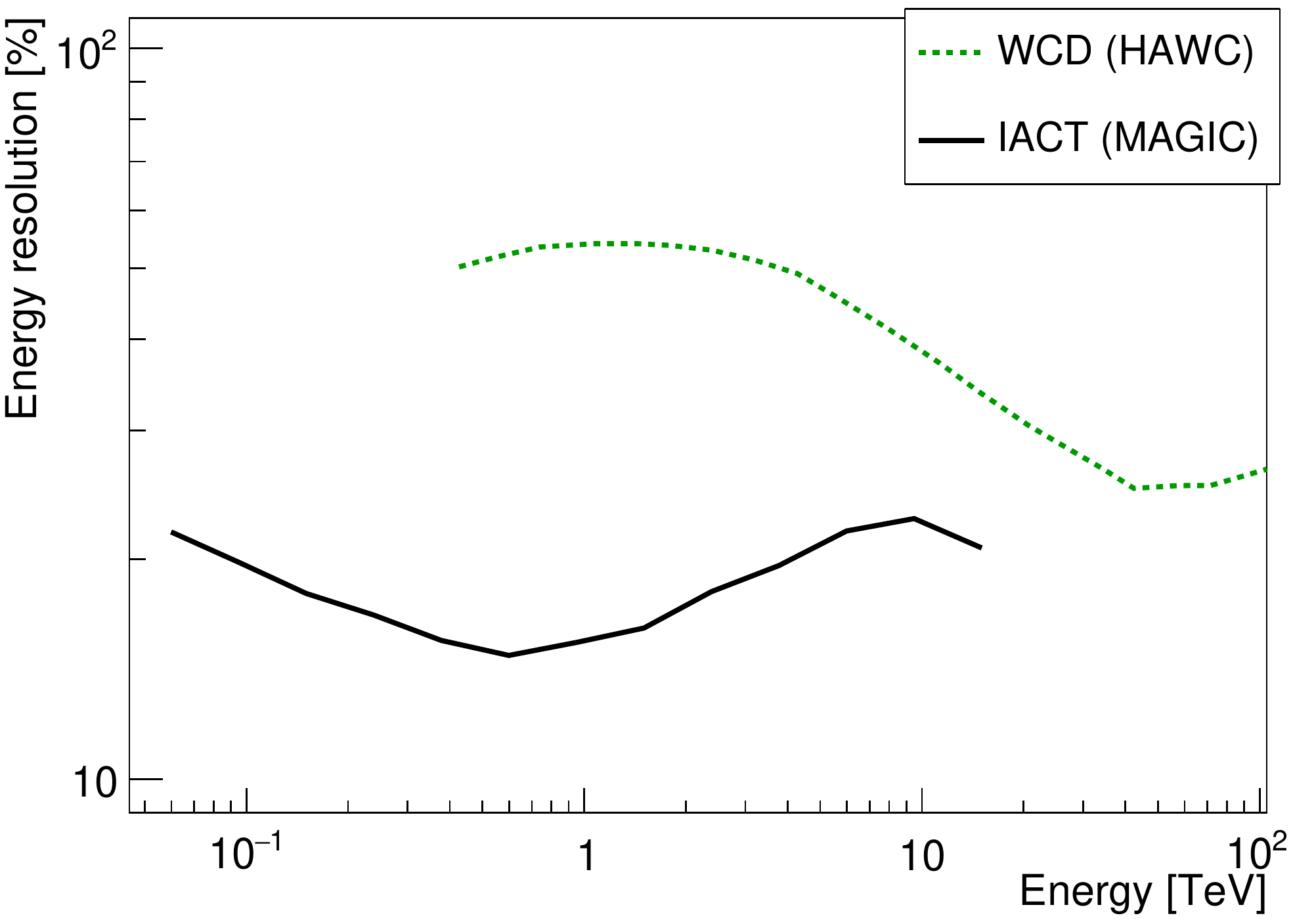}
    \caption{Comparison of the angular (left panel) and energy (right panel) resolution of WCD (HAWC, dotted green line) and IACT (MAGIC, using look-up tables energy estimation, solid black line). 
    HAWC angular resolution is taken from \url{https://www.hawc-observatory.org/observatory/sensi.php}, and energy resolution (for neural networks analysis, recalculated from $\log_{10}$ space to linear space) from \cite{2019ApJ...881..134A}. MAGIC performance in both panels is taken from \cite{2016APh....72...76A}.}
    \label{fig:ares_eres}
\end{figure}

In Table~\ref{tab1} the basic performance parameters are summarised and compared for IACT and WCD/SA detectors.
\begin{specialtable}[t]
\caption{Comparison of the typical performance parameters of IACT and SA/WCD instruments. Due to energy dependence of the performance parameters, the energy resolution, angular resolution and sensitivity are given for the best sensitivity range of IACT or SA/WCD instruments (i.e. about an order of magnitude higher than their corresponding energy thresholds).
The values for individual instruments of a given type can vary by a factor of a few depending on the details of the hardware implementation and analysis methods. \label{tab1}}
\begin{tabular}{p{0.29\linewidth}p{0.32\linewidth}p{0.30\linewidth}}
\toprule
\textbf{Characteristic}	& \textbf{IACT}	& \textbf{SA/WCD}\\
\midrule
Energy threshold & $\sim$ tens of GeV (for a few hundred m$^2$ mirror dish) & $\sim$TeV \\
Duty cycle & $\sim10\%$ & $\lesssim100\%$ \\
Field of view & $\sim$ a few millisr & $\sim$ sr\\
Energy resolution & $\sim15\%$ & $\sim 40\%$ \\
Angular resolution & $\sim 0.1^\circ$ & $\sim 0.2^\circ$ \\
Sensitivity & $\sim1\%$ Crab Nebula flux in 25\,hr& a few \% Crab Nebula flux in 5\,yr \\\hline
Main present instruments & \mbox{H.E.S.S., MAGIC}, \mbox{VERITAS} & Tibet AS-$\gamma$, HAWC, LHAASO-WCDA, LHAASO-KM2A\\
Future instruments & CTA & SWGO, ALPACA \\
\bottomrule
\end{tabular}
\end{specialtable}
The energy dependence of the angular and the energy resolution for IACT and WCD instruments is shown in Fig.~\ref{fig:ares_eres}.
The performance parameters of both types of instruments in general improve with energy, however at the highest energies they might again worsen due to saturation effects.
The maximum energy up to which the source can be studies is strongly dependent on both the geometrical footprint of the instrument (favouring sparse arrays), but also on the spectral shape and flux of the observed source. 

IACTs detect the shower development already in the atmosphere rather than the tail of the shower that reaches the ground.
This has a profound impact on the energy threshold and the low-energy (sub-TeV) performance of those instruments.
Such low-energy showers will develop mainly in the atmosphere, providing an ample information  and only a small number of particles will reach the ground, providing little input for WCD and SA techniques. 
Sparse SA are sensitive only above about 10\,TeV. 
Full-coverage approach used in some WCD and e.g. ARGO-YBJ allows observations starting from (sub-)TeV. 
On the other hand WCD and SA instruments are not limited to observations during (preferentially moon-less) nights, and in addition have a large field of view (FOV), thus can monitor continuously a large fraction of the sky. 
This makes also difficulty in comparison of sensitivity between the two types of instruments. 
While for transient phenomena and short-term variability IACTs are clearly superior, the possibility of integration of a large exposure from WCD or SA instrument from broad FOV is an efficient method both for steady sources studies, and for unbiased studies of typical states of variable emitters. 
Those differences are reflected in the exposure time quoted for sensitivity calculation. 
In the case of WCD and SA the sensitivities are typically quoted for 1 or 5 years, and they are related to the total, aggregated time of instrument operation.   
In contrary, for IACTs 50\,hrs are typically used as a reference number. 
While it is theoretically possible to observe a single source with an IACT even for 300\,hr within a single season, in practice it is usually not done as the sensitivity in (sub-)TeV range improves only with square root of time, and extensive exposures would block observation time of other targets.  
In addition, at the lowest energies the sensitivity starts to be limited by the systematic uncertainties on the background, preventing further improvement with increasing observation time.  

As IACTs probe via imaging technique, a significant fraction of the shower developments (in particular around the shower maximum), they are fairly good at calorimetry of the parent particle. 
The achieved energy resolution of IACTs in general improves with energy, and depends strongly on the energy estimation method, however values of the order of 15\% are typical. 
On the other hand the number and total energy of particles in the tail of the shower, that are accessible to the WCD and SA techniques is subjected to high shower-to-shower fluctuations.
This effect strongly degrades the energy resolution of the WCD and SA instruments. 

IACT and WCD instruments both achieve angular resolution of the order of $0.1^\circ$ in their highest  energy ranges.
However, as the energy range is different for both types of instruments at the low (TeV and sub-TeV) energies IACT clearly provide higher performance.
At energies of tens of TeV both types of instruments have excellent angular resolution.
However, WCD/SA excel by higher duty cycle, which improves the sensitivity and thus facilitates investigation of source morphology. 

Because of a much different duty cycle of both types of instruments, and their instantaneous performance it is difficult to fairly compare them in terms of sensitivity. 
The same sensitivity of about $3\%$ of Crab Nebula flux in a given sky location is achieved by  WCD in about 5 years and by IACT in just $\sim$3\,hrs. 
However WCD would simultaneously scan about half of the sky with such a sensitivity.
To cover the same fraction of the sky with such a sensitivity IACTs would require about 8000\,hrs of observation time (assuming 2 msr FoV), which corresponds to 8 years of observations with a typical duty cycle of IACTs.  
Therefore large-scale unbiased sky surveys with IACTs are feasible only for regions in which an enhanced number of sources is expected, such as the Galactic plane \citep{2018A&A...612A...1H}
or the Cygnus region \citep{2018ApJ...861..134A}.
On the other hand WCD turned to be excellent survey instruments, in particular for the highest energy Galactic sources \citep{2017ApJ...843...40A, 2021Natur.594...33C}.

Large FOV instruments can serve also as a source of trigger of more detailed observations with a high-sensitivity, pointed instruments.
This synergy can be used between WCD/SA and IACT either to study stable sources discovered with deep WCD/SA observations or to follow up short-term flares. 
VERITAS and MAGIC telescopes performed follow-up observations of the sources discovered in the HAWC catalogs \citep{2018ApJ...866...24A,2019MNRAS.485..356A}. 
Surprisingly, those studies did not confirm TeV emission from most of the reported sources.
Likely explanation of this apparent tension between results of both techniques is given by possible $1^\circ$  scale extension of these sources.
At energy of $\sim$\,TeV such an extension is comparable to angular resolution of HAWC (and thus would also not affect its sensitivity strongly), while it would degrade the sensitivity of IACT observations by about an order of magnitude, and in more extreme causes could cause confusion of the source emission with the isotropic background.   
In fact, a joint study of H.E.S.S. and HAWC was performed in which IACT data were artificially smeared to match the WCD angular resolution and a similar method for background subtraction was used for both types of instruments.
It showed that both techniques provide consistent view of the Galactic plane with remaining differences in the gamma-ray flux explainable by the systematic uncertainties \citep{2021ApJ...917....6A}. 

The second possibility, triggering IACTs on short-term flares seen by WCD/SA has been so far much less explored. 
The main reason is that short-term sensitivity and limitations from the false alarm rate over a broad FoV.
This makes detection efficient only for very strong flares, larger than a few times the flux of the Crab Nebula \cite{2017ApJ...843..116A}, which have been observed only in the handful of brightest blazars. 

\subsection{Hybrid arrays}

The synergies of both types of techniques allow to combine them within the same facility. 
This has been done already in the previous generation of instruments with combination of HEGRA IACTs, wide-acceptance Cherenkov array AIROBICC and surface detectors (scintillator arrays) \citep{2002A&A...390...39A}.

Modern version of such a combination is LHAASO, a hybrid array combining various TeV observations techniques to achieve excellent performance over broad energy range both for observations gamma rays and cosmic rays \cite{2010ChPhC..34..249C}.
The facility has been completed in 2021 and it is starting to provide first scientific results (see e.g. \citep{2021Natur.594...33C}). 
It composes among others: LHAASO-KM2A -- a sparse array of electromagnetic surface detectors (scinitilation counters) and underground WCD for muon detection, LHAASO-WCDA (Water Cherenkov Detector Array) -- a three tessellated water ponds, and LHAASO-WFCTA (Wide Field air Cherenkov Telescopes array) -- an array of 18 IACT with a wide FOV ($16^\circ$ x $16^\circ$). 

\section{Future instruments}

In the following years the next generation of TeV instruments will be constructed.
In the case of IACT instruments, a major leap in performance is expected between the current generation and the Cherenkov Telescope Array (CTA,  \citealp{2013APh....43....3A,2019scta.book.....C}). 
CTA will be composed of about a hundred of telescopes in three different sizes (LST, MST and SST).
In order to provide all-sky coverage, the telescopes will be distributed into two locations (one in the Northern and one in the Southern hemisphere). 
The arrays of four LST (Large-Sized Telescope) due to their 23-m diameter reflectors will allow one to firmly extend the IACT technique into the still poorly exploited region of tens of GeV.  
SST (Small-Sized Telescopes) are large (70 units) arrays of a few-metre-diameter telescopes. 
They  are aimed at providing optimal sensitivity, angular resolution and surveying capabilities at energies of tens of TeV \cite{2020A&A...634A..22L}, making them competitive instruments with WCD/SA technique.
The necessity of large FOV for SSTs without introducing large optical aberration stimulates usage of previously unexplored or weakly explored technologies for IACTs, such as dual-mirror (Schwarzschild-Couder) design and usage of SiPM light detectors.
MST (Medium-Sized Telescopes) are the most similar in diameter to the current generation of IACT.
Nevertheless, their large number (15 and 25 in Northern and Southern CTA site) will allow improvement of the sensitivity in the canonical energy range of IACTs ($\sim$TeV) by an order of magnitude, and large FOV will allow one to perform sensitive scans of large parts of the sky. 
Two different designs are pursued for MSTs: in addition to the classical single mirror also a dual-mirror design \citep{2021APh...12802562A}.

The sensitivity of WCD instruments is mainly in TeV range. 
In the case of distant extragalactic sources the TeV emission is severely absorbed in the pair production process on extragalactic background light \citep{1967PhRv..155.1408G}.
Capability of studying sources with the redshift $\gtrsim0.1$ with WCD instruments is therefore strongly limited \citep{2021ApJ...907...67A}.
On the other hand, Galactic sources often show broadband spectra extending into tens of TeV (and beyond) regime. 
Thus it is unfortunate that all currently operating WCD and SA installations are located in the Northern hemisphere (Tibet AS-$\gamma$ at $30^\circ$N, HAWC at $19^\circ$N,  LHAASO at $29^\circ$N), from which location most of the Galactic plane is not observable. 
Southern Wide-field Gamma-ray Observatory (SWGO, \citealp{2019arXiv190208429A}) is a project of building a WCD experiment in the Southern hemisphere.
The project is currently at an early design phase with various detector concepts \citep{Werner:2021d/}. 

Another planned project that will exploit surface array of detectors for gamma-ray observations in the Southern hemisphere is ALPACA and its prototype array ALPAQUITA \citep{2021ExA....52...85K}. 
ALPACA will have a similar design to the Tibet AS$\gamma$ experiment, with a sparse array of scintillation detectors and underground muon detectors.  

\begin{figure}[t]
    \centering
    \includegraphics[width=0.65\textwidth]{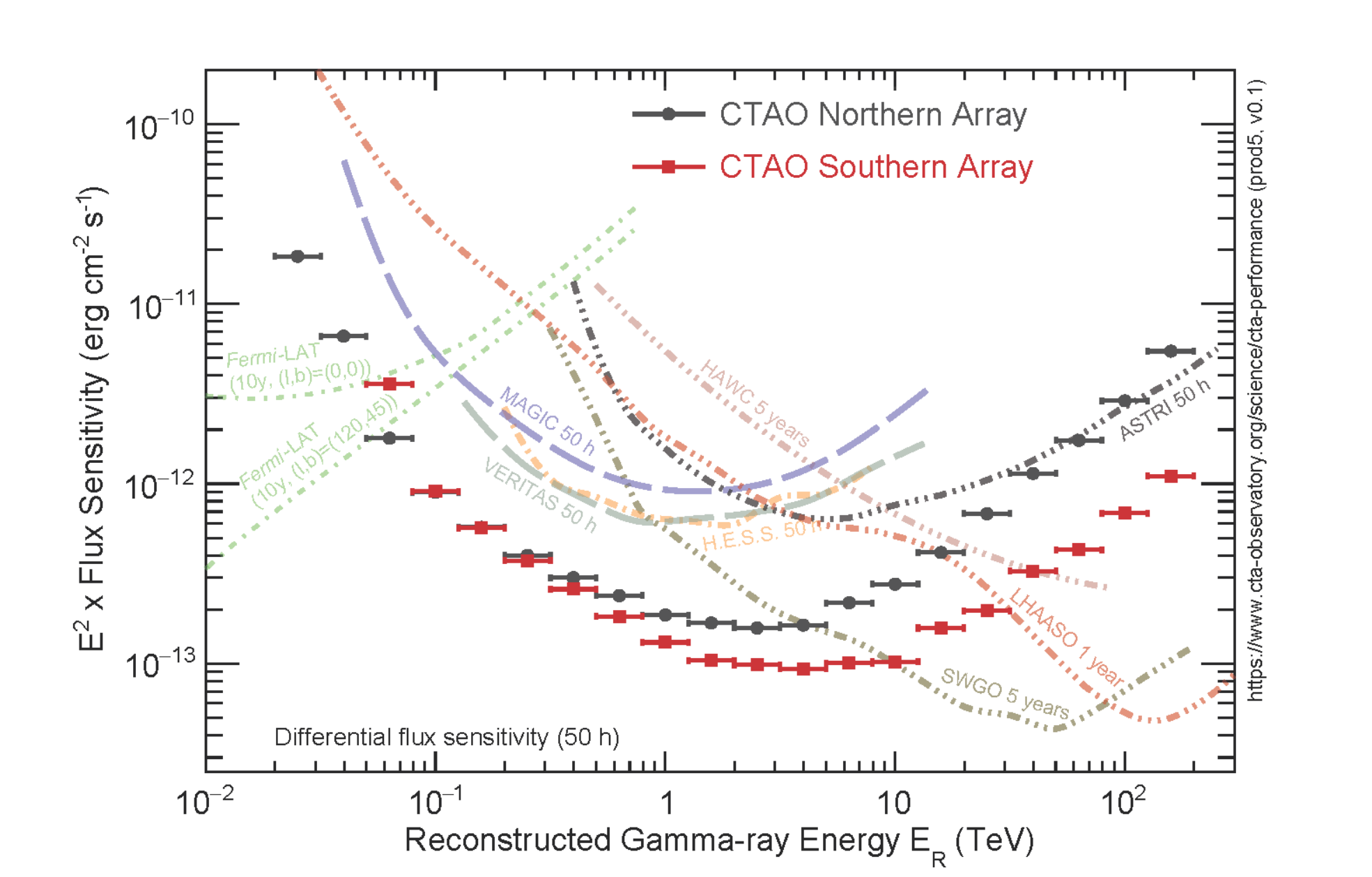}
    \caption{Comparison of sensitivity of present and future GeV and TeV instrumentations:
    ASTRI \citep{Lombardi:20214O},
    CTA \citep{cherenkov_telescope_array_observatory_2021_5499840},
    \textit{Fermi}-LAT  (\url{http://www.slac.stanford.edu/exp/glast/groups/canda/lat_Performance.htm}, scaled),
    HAWC (\citealp{2017ApJ...843...39A}, scaled), 
    H.E.S.S. (adapted from \citealp{2015arXiv150902902H}),
    MAGIC \citep{2016APh....72...76A},
    LHAASO \citep{2019arXiv190502773B},
    SWGO \citep{2019arXiv190208429A},
    VERITAS (\url{https://veritas.sao.arizona.edu/about-veritas/veritas-specifications}).
    Image reproduced from \url{https://www.cta-observatory.org/science/ctao-performance/}.
    }
    \label{fig:sens_comp}
\end{figure}

The comparison of sensitivity of various presently operating and future instruments is shown in Fig.~\ref{fig:sens_comp}.
The construction of the next generation instruments will allow us to perform studies in poorly investigated so far energy ranges (tens of GeV and above tens of TeV) and significantly improve the sensitivity in TeV energies. 
This is expected to turn detections of individual sources of rare classes (such as gamma-ray bursts, novae, and some sub-classes of active galaxies)  up to the level of population studies.
Discoveries of gamma-ray emission from previously unknown in this energy range classes of objects is also likely to be expected. 
Finally, the survey capabilities of those instruments will allow unbiased search for gamma-ray emission. 

\section{Analysis methods}

The analysis of TeV data, both from IACT and WCD/SA instruments can be divided into three stages.
Low-level analysis takes care of calibration (both in terms of reconstructed signal strength and its timing) of individual photomultipliers. 
In the medium-level analysis the information from the individual detectors is aggregated and based on it the basic parameters of a single event (such as event type, its energy and arrival direction) are determined. 
In the case of WCD and SA, instead of physical shower parameters sometimes more experimentally-oriented quantities are used at this step (e.g. instead of event energy, the fraction of tanks hit by the shower). 
In the last analysis stage, the individual event information are used to derive high-level information about the source (such as the energy spectrum, morphology of the emission, or time variability). 

As both WCD/SA and IACT instruments share the same principle of reconstruction of air showers, some of their analysis problems are shared as well.
In particular, gamma-ray sources are immersed in a few order of magnitude more abundant isotropic background induced by cosmic ray showers.  
The sensitivity of TeV instruments is mainly dependent on how well such a background can be rejected and how large systematic uncertainties are induced by the subtraction of the residual background. 

\subsection{Event cleaning}
In both techniques low-energy showers, that are actually the most abundant in the spectra of cosmic sources, produce smaller response in the instrument, more affected by the noise. 
This results in only a small fraction of pixels (in IACT), tanks/detector stations (in WCD and SA respectively) carrying the shower information.
For such events it is essential for achieving low energy threshold and good low-energy performance to determine which parts of the detector are dominated by the shower rather than by the noise.  

In the case of IACT the brightest pixels in the camera  (above a given threshold) form the core of the image. 
Additional condition is used to accept also lower brightness pixels if they are neighbouring one of the core pixels \citep{1991ICRC....1..464P}.
Usage of the signal timing significantly improves the reconstruction of low-energy showers \citep{2009APh....30..293A,2013ICRC...33.3000S}.

In the case of WCD, the events in which the number of triggering tanks is too low are excluded from the analysis, and the events are classified based on how big fraction of all the tanks provides a trigger \citep{2017ApJ...843...39A}. 

\subsection{Event reconstruction and background rejection}

In the case of IACTs the most commonly used background rejection is still based on the classical Hillas parametrization of the events \citep{1985ICRC....3..445H}.
Due to the intrinsic differences of gamma-ray and hadronic induced air showers, the latter ones tend to be more spread out, in particular in the lateral direction. 
Additionally stereoscopic observations of the same event with multiple IACTs allow 3D reconstruction of the event, including the estimation of the height of the peak of the Cherenkov emission (which is a measure of the maximum of the shower development).
This allows an efficient rejection of single muon events that could mimick gamma-ray showers (see e.g. \citealp{2012APh....35..435A}) and for a partial rejection of hadronic events dominated by a single electromagnetic subcascade \citep{2007APh....28...72M, 2018APh....97....1S}.

In the case of WCD and SA the reconstruction of the shower parameters is obtained from the distribution of deposited energy or particle density at the ground level. 
The lateral distribution of particles in the shower is described by Nishimura-Kamata-Greisen function \cite{1960ARNPS..10...63G}.
Therefore, the reconstructed signals on the ground can be fit as a function of the distance from the core \citep{2017ExA....44....1K, 2017ApJ...843...39A}. 
In HAWC, hadron-induced events are excluded based on the clumpiness of the lateral distribution (sporadic high signals at large distances from the core position, caused by muons and hadronic subshowers), \cite{2013APh....50...26A}. 
Another parameter that is used is the deviation from the axial symmetry of the lateral distribution \cite{2017ApJ...843...39A}.
Optimal values of those cuts can be then selected depending on the extend of the event in the detector (closely related to its energy). 
SA with dedicated, underground detectors, such as Tibet AS-$\gamma$ and LHAASO-KM2A, can directly count the number of penetrative particles (measuring mainly muons) exploit this information  for gamma/hadron separation (see e.g. \cite{2021ChPhC..45b5002A}). 
This allows for background-free regime above about 100\,TeV. 
Also the shower age parameter can be used to discriminate gamma rays from background events (see e.g. \cite{2011ChPhC..35..153F}).

Machine learning methods using tree classifiers (such as Random Forest or Boosted Decision Trees) are commonly used both in IACT \cite{2008NIMPA.588..424A, 2009APh....31..383O, 2017APh....89....1K} and WCD \citep{2021arXiv210800112C} data analysis to aggregate the information from different shower/image parameters into a single gamma/hadron separation parameter.

The arrival direction reconstruction in WCD technique is mainly based on arrival time distribution along the instrument.
This can be exploited with a planar fit \cite{2003ApJ...595..803A}.
Additional conical correction and iterative procedures can be used to further improve the direction reconstruction.
For point-like sources, the angular resolution improvement also lessens the effect of the isotropic background as it is integrated over a smaller solid angle around the position of the source. 

Another method for reconstruction of IACT events exploits comparison of shower images with expected images either from a semi-analytic model \citep{1998NIMPA.416..425L,2009APh....32..231D} or by using MC-generate templates \citep{2014APh....56...26P}.
Similarly, MC-template-based analysis has been also applied to WCD data \citep{2019JCAP...01..012J}. 
Fitting of the individual events to such model provide direct information about the physical properties of the shower (such as arrival direction, energy, impact, height of the first interaction). 
Goodness of the fit can be used for gamma/hadron discrimination, however it requires rather precise knowledge of the underlying probability distributions of the pixel response to gamma-ray showers (including also noise), hence it is prone to systematic errors \citep{2011APh....34..858B}.   
Hillas-based parameters of IACT technique can be also combined with the parameters obtained from the shower model for an improved performance using multivariate methods \citep{2010APh....34...25F}. 

\subsection{Background modeling}

Even after gamma/hadron separation TeV instruments still need to deal with the residual background\footnote{Only at the highest energies for strong sources the observations can be considered background-free.}.
It is essential for the background estimation to be determined precisely, because even few per cent discrepancy can produce significant systematic error on the measurement of source flux measurement, possibly even causing an artificial signal. 
Due to this, for IACT's sensitivity definition, the expected excess must not only achieve $5\sigma$ statistical significance, but also be above $\sim5\%$ of the residual background. 
To limit the systematic errors affecting the background estimation, the observation conditions, such as zenith angle, and atmosphere transparency, needs to be as similar as possible for the ``source'' and ``background'' data.  
The simplest method to estimate the background for point-like sources is an analogue of the aperture photometry. 
In the case of pointed instruments such as IACTs this can be realised by tracking a false source, at a given offset from the actual target \cite{1994APh.....2..137F}.
The source position will then slowly rotate around the camera centre.
Its reflected position (one or more) with respect to the centre of the camera can be used as a background control region.
To counteract the camera inhomogeneities the position of source and anti-source is swapped every few tens of minutes, the so-called wobbling. 
Thanks to this method background can be estimated down to systematic accuracy of $\sim1\%$ (see e.g. \citealp{2016APh....72...76A}).
Noteworthy, such swapping of source and background positions will be only effective against instrumental inhomogeneities and not the ones connected with the given sky position (such as stars or  nearby, possibly extended TeV sources). 
In particular, the presence of bright stars within $\sim1^\circ$ from the source position might require a dedicated correction procedures \cite{2011JCAP...06..035A}, such as simulating the effect of the star in the background.

With the total number of detected sources at TeV energies increasing, nearby TeV emitters are more and more prone to additionally complicate the background subtraction. 
This is especially problematic in the case of crowded regions, such as the Galactic plane \citep{2018A&A...612A...1H}. 
On one hand, such sources should be taken into account when background estimation regions are selected (see e.g. \cite{2012A&A...544A.142A}).
In the case of scans with IACTs, as long as the inhomogeneity can be kept low, the background can be estimated from a ring around the source position, excluding parts of the ring that can be affected by other sources \cite{2018A&A...612A...1H}, the so-called adaptive ring background method.
On the other hand the emission of nearby sources (see e.g. \citealp{2016ApJ...821..129A}) can overlap, either by actual coexistence in a particular direction of the sky of extended sources, or due to the smearing of the observed emission by the gamma-ray PSF of the instrument.  
Let us consider an example of two nearby sources: A and B.
To obtain the flux of A, one needs to subtract the contribution of the B source, which depends on the flux of B.
However to obtain the flux of B, similarly one would need to know the flux of A. 
This can be solved by simultaneous likelihood fits of both sources with particular morphology and spectral models \cite{2018A&A...619A...7V}.
Similar method is commonly used by GeV satellite instruments (see e.g. \citealp{2010ApJS..188..405A}).

In the case of WCD and SA different algorithms can be used to estimate the isotropic background. 
The equal-zenith method \citep{1999ApJ...525L..93A} is similar to aperture photometry, with the background taken from multiple OFF regions adjacent to the source. 
Due to strong zenith angle dependence of the event rate, the background estimation regions are selected preferentially at the same instantaneous zenith angle as the source.  
The second method, direct integration algorithm \citep{2012ApJ...750...63A} exploits decomposition of the background into time-independent angular distribution (expressed in local coordinates) and direction-independence of the time-dependent total rate of events.
This method is accurate down to a fraction of per mil, limited by small cosmic ray anisotropies \citep{2012ApJ...750...63A}.
Also in the case of WCD, likelihood analysis was used to fit source morphology and take into account partially overlapping sources \citep{Huang:2021bE}.

\subsection{Largely extended/diffuse emission}

Spatial extension of TeV emission can be challenging, in particular to IACT instruments. 
Firstly, because the underlying background is then larger, considerably worsening the sensitivity.
Moreover, very extended sources, with a radius of $\gtrsim1^\circ$, have its extend comparable to IACT cameras. 
This results in the lack of background control regions, making analysis very difficult\footnote{If the extension of the region is mainly in one direction, such as inner part of the galactic plane\cite{2018A&A...612A...9H}, it is still possible to use adaptive ring method to evaluate the background. }. 
WCD and SA detectors, as wide FOV instruments are more suitable for such sources. 

A specific case of extended sources is the all-sky diffuse emission. 
One example of it, is the measurement of diffuse electron contribution to cosmic rays \citep{2009A&A...508..561A}. 
In such a case there is no possibility to estimate the background from the data itself. 
Instead MC simulations of protons are used, and normalised to the data using proton-like events. 
In such studies a careful selection of data quality and hadron rejection cuts is essential, as on one hand the background needs to be kept as low as possible to minimise the systematic errors, but on the other hand, the studies are also sensitive to data/MC mismatches that are often more pronounced for strong selection cuts. 
The diffuse emission of the background itself can be also studied to derive the cosmic ray spectrum\cite{1999PhRvD..59i2003A,Temnikov:2021Iq}.
Interestingly, this type of study is limited by the need of extensive MC generation rather than by the data collection itself. 

In contrast, the wide FOV of WCD/SA instruments, combined with the background direct integration method, allow detection of sources with an extend of a few degrees \cite{2017Sci...358..911A}.
The WCD and SA experiments are also excellent instruments for studying (diffuse) cosmic rays \citep{2019ApJ...871...96A}. 
Large structures, such as Galactic disk, can be efficiently studied with the SA/WCD techniques as long as background control regions can be established \citep{2021PhRvL.126n1101A}.
Nevertheless, for all-sky fully diffuse emission on top of the cosmic-ray background, they struggle with the same background estimation issue as IACTs.

\subsection{Energy spectrum}
Due to both the intrinsic shower fluctuations (see e.g. \citealp{1998APh.....9...45C}) and the measurement accuracy, there is a considerably energy dispersion. 
This effect can cause unbalanced migration of events between the considered energy bins, particularly for spectra that are either soft, or have a cut-off. 
The conversion from estimated energies to true energies can be done by means of unfolding procedure (see e.g. \citealp{2007NIMPA.583..494A} and references therein). 
While the problem naively can be seen as a simple inversion of an energy migration matrix, the typical form of this matrix makes the inversion strongly dependent on even tiny changes of the input matrix. 
This can lead to nonphysical, oscillating solutions. 
There is no unique way how to deal with this problem and different approaches might be more appropriate for individual cases. 
Most of the methods involve a free parameter (regularization) that describes the balance between smoothness of the obtained solution, and its accuracy in description of the data. 

A special method of unfolding is the ``forward folding''.
Namely, a given source spectrum (expressed in true energy) is folded with the instrument response function into measurable quantities (typically the estimated energy).
Comparison with the data can be performed e.g. by likelihood maximisation. 
This method is in general more reliable than regular unfolding, in particular in the case of a strong energy migration. 
It can be also easily generalised to multiple or extended sources by performing a joint likelihood fit in energy and direction. 
Both reasons make it a natural method to be applied for wide FOV instruments (see e.g. \citep{2017ApJ...843...39A} for application to WCD), however they have been also used for IACT (see e.g. \citealp{2018A&A...619A...7V}).
A disadvantage of such a likelihood fit is a need for assumption of a given source spectral form. 
Therefore the result of the fit is only a set of spectral parameters, rather than spectral measurement at different energy ranges. 
Typically simple phenomenological spectral shapes (power-law, log-parabola, power-law with a cut-off, etc.) are used.
While for a strong source complicated spectral shapes can be fit, low-significant sources are usually only possible to be fit with the simplest (i.e. having least number of free parameters) spectral shape of power-law. 
This neglects any intrinsic curvature of the spectra, and in some cases can bias the derived results from such studies, especially if they are made over a population of sources (namely, while the curvature is not significant in individual source, it is apparent in the combined analysis of multiple sources, see e.g. discussion in \citealp{2019MNRAS.486.4233A}). 
It should be noted that even if the source spectrum at a given moment of time might follow closely a simple shape (e.g. a power-law), spectral variability can cause the effective spectrum integrated over an extended period of time to be quite complicated. 
Such complicated shapes cannot be parametrised easily, or the number of parameters is too large for efficient fitting. 

\subsection{Deep learning methods}

The standard analysis approach of both IACTs and WCD involves distillation of information from individual waveforms into pixels/tanks information and further into a few parameters characterising a given event.
At each step assumptions and simplifications are done that potentially lose pieces of information and hence worsen the achieved performance of the method. 
An alternative, assumption-free approach is to provide as full information about each event as possible, and employ a machine learning method to find the best solution, the so-called deep learning (DL, \citealp{goodfellow2016deep}). 
Such techniques can be used in all the stages of event reconstruction: for gamma/hadron separation, energy estimation or arrival direction estimation. 
While the idea is not new (see e.g. application of neural networks classifier to Whipple telescope data, \citealp{1995APh.....3..137R}) it raised rapidly in popularity in the recent years, both due to the increase of the computer power and also due to development of publicly-available algorithms for classification of optical images. 
A natural way of tackling the problem of reconstruction of events based on its pixelized images is the usage of convolutional neural networks (CNN). 
This method is exploiting a sequential usage of filters for detection of the image features and pooling layers that condensate the information. 
On the other hand recurrent neural networks (RNN) are designed to handle data series with varying length by internal memory of the inputs to exploit the correlation between sequential inputs. 
While their architecture is not directly suited for analysis of 2-dimensional events, their combination with CNN has been shown to be an efficient way of combining information from multiple IACT telescopes \citep{2019APh...105...44S,2020EPJC...80..363P}.

There are a few challenges in application of DL methods to TeV data. 
First, and the most important, such methods are very sensitive to any differences between the two provided training sets. 
The reason is that the information about the shower is hidden in a much more abundant pixels, tanks or detector stations that carry only noise information. 
For example if a gamma/hadron separation is trained on MC gamma-ray events and real data hadron events a DL method might be more prone to focus on the small MC/data discrepancies rather than on the actual difference between gammas and hadrons \cite{2019APh...105...44S}. 
Training on gamma-ray and hadron samples produced solely from MC simulations is possible but not always feasible. 
Namely, hadronic showers are more CPU-time-consuming to generate, the samples are usually not complete (both due to possibility of triggering small showers at large impact parameters\footnote{in the case of IACTs also at large offset angles from the camera centre}, and due to the distribution of different nuclei present in the cosmic spectra).
Additionally, contrary to the TeV gamma-ray MC simulations that are nearly exclusively using well established electromagnetic process, the accuracy of the cosmic-ray MC simulations suffer from the differences between available hadronic interaction models \cite{2021JPhG...48g5201O}.
Moreover, when the training is performed using only MC simulations, application of it to the real data might show performance losses.  
In contrary, training on data-only is limited by the need of proper gamma-ray samples -- even from the direction of strong sources, gamma rays are only a small fraction of the total number of events, hence preselection of particle type is needed.
Also in this case crucial information (such as the true energy of primary particle) is not available. 

Second, efficient usage of DL methods requires large samples (the larger, the more parameters are exploited), which in turn results in very long processing times. 
This has been mediated in the recent years by usage of graphical processor units (see e.g.~\citealp{NietoCastano:2017/K}). 
Third, most of the generally available DL methods used for image analysis operates on square pixalisation, that is common in regular images. 
In contrast, the TeV instrumentation is often using hexagonal pixel/tank configuration.
Interpolation methods can be used to translate the events into square pixelisation. 
However, dedicated DL methods using intrinsically hexagonal pixels are also available \cite{2019SoftX...9..193S}.

In the last few years a large progress in the application of DL methods to IACT instruments has been observed. 
While the first DL results performed with MC simulations were showing in the past (sometimes large) improvements, when the training was applied to the actual data the DL methods often turned out to perform worse. 
Nowadays there are multiple reports of confirming improvements of DL-based analysis also with real IACT data, however still the obtained improvement is smaller than expected from MC-only studies. 
In the case of H.E.S.S. telescopes data, DL methods were shown to provide a boost in the gamma/hadron separation of actual data, however slight worsening in the angular resolution has been observed as well \cite{2019APh...105...44S}. 
Similarly, in the case of LST-1 telescope commissioning data, DL methods provided an improvement in the gamma/hadron separation, but not in the angular resolution \cite{2021arXiv210804130V}.
CNN method applied to cleaned images of showers in MAGIC data resulted in similar sensitivity to the standard approach based on decision trees \cite{2021arXiv211201828M}. 

The DL methods can go one step deeper and in the case of pixel-wise information, use the whole, sampled PMT waveforms instead of single charge/time measurements per pixel \citep{2021APh...12902579S}. 
A MC proof-of-concept study showed that this can further improve the gamma/hadron separation for IACT. 
However, this latest improvement still remains to be confirmed with an actual IACT data. 

The deep learning methods are also starting to be exploited for WCD and SA instruments.
While machine learning methods are shown to improve the performance of the gamma/hadron separation, they have been fed so far just with a small selection of event parameters, rather than the full event information\cite{2021arXiv210800112C}.
Treating of HAWC events as images in CNN has been tried as well, and has shown a gamma/hadron separation capability \citep{Watson:2021K+}, however it is not clear yet if such a method improves the performance.  
It is likely that as the understanding of these experiments improve in time, more complete information can be exploited as well. 
Usage of deep learning for LHAASO-KM2A using MC simulations shown improvement in the gamma/hadron separation with respect to the standard method based on the ratio of the detected muons (by underground detectors) and electrons/positrons (by surface detectors) \citep{Zhang:2021ki}, however the method has not been validated on the data yet.  

\subsection{Combination of data from different instruments}

Multiple TeV instruments are currently in operation.
Often the information on TeV emission of a given source is available from more than one instrument, either by observing independently, responding to the same Target of Opportunity (TOO), or by pre-planned, joint observational campaigns. 
While high-level products, such as light curves or spectra might be sometimes combined using statistical methods, this is not always feasible due to e.g. different time or energy binning used. 
Moreover as the actual spectral points generally depend on the best fit spectral shape, inclusion of additional information from one experiment would change also the spectrum obtained by the other.   
Finally, in the case of non-detection of gamma-ray emission, combining upper limits from individual instruments is only possible if additional information (such as likelihood profiles, see e.g.  \citep{2021arXiv210813646A}) is available. 
Joint analysis of data from different instruments is marred by the fact that different (and sometimes proprietary) software and data formats are used by different collaborations.
Nevertheless, in the recent years a lot of progress was done in terms of unifying the TeV astronomy data format \cite{2021Univ....7..374N}.
This allowed first common analysis of data from different instruments performed within one framework \cite{2019A&A...625A..10N}.
Interestingly, that analysis involved not only WCD and IACT instruments, but also GeV satellite data. 
It should be noted however that for strong sources the statistical uncertainties of the individual measurements are generally much smaller than systematic differences between the instruments (see e.g. \citealp{2020ApJS..248...29A}).  

\section{Multiwavelength and multi-messenger observations}

In order to better understand the processes governing cosmic sources a broadband view is essential.
The observations can be either organised as multiwavelength (MWL), if TeV band is combined with other electromagnetic observations (e.g. radio, optical, X-ray) or multi-messenger (MM) if gamma-ray observations are combined with studies of e.g. neutrino emission, or gravitational waves. 
The common observations can be performed either in terms of pre-planned multi-instrument campaigns or as a response to a TOO. 

In the case of TOO observations, WCD and SA as wide FOV instruments have an advantage of being able to measure (if source is above the instrument's horizon) the source emission also before and during the TOO time. 
However, the performance of those instruments over short time scales is relatively poor, hence only extremely strong emission can be probed this way.
On the other hand IACT, if designed for this goal and equipped with appropriate software for automatic reaction, are able to repoint to any position in the visible sky within a timescale of few tens of seconds (see e.g. \citealp{2016APh....72...61A}).
It should be noted however that such immediate IACT observations might turn out in non-optimal conditions (such as high zenith distance angle, bad weather conditions, or moonlight, all of them increasing the energy threshold of the observations). 

\subsection{TOO observations with a large position uncertainty}
Particularly difficult case is the follow-up observations of single events which cannot be associated firmly to a known source and thus have a large position uncertainty.
The positional accuracy of MWL triggers such as fast radio bursts \citep{2021arXiv210604352T} or gamma-ray bursts \citep{2009MNRAS.397.1177E} normally depends on the electromagnetic flux of the event and the instrument from which the alert originated. 
The uncertainty of the position thus typically vary between much less than IACT PSF and a fraction of a degree.  

The localisation accuracy is usually worse in the follow-up of MM triggers. 
In particular, the current localisation accuracy of the gravitational wave events \citep{2019CQGra..36n3001B} can easily cover tens of square degrees region in the sky, significantly larger than the FOV of IACTs.
This requires a clever pointing strategy that maximises the chance of detection of the electromagnetic counterpart (see e.g. \citealp{Ashkar:2021Hu}).  
In the case of a follow-up of a single high-energy neutrino event \citep{2016JInst..1111009I}, the localisation accuracy is usually much better (of the order of a fraction of degree), which is however still considerably larger than the PSF of IACTs.

This has a direct effect on the detection capability of such events. 
For a source with an unknown location, it is not sufficient to achieve detection of an excess from some location within the error circle of the neutrino localisation with a $5\sigma$ statistical significance, because the number of trials associated with different possible locations of the source has to be taken into account.
As an example, if the localisation uncertainty is $\sim1^\circ$ and the PSF is $\sim0.1^\circ$, the number of trials can be roughly estimated as $\sim 100$.
In such a situation in order to achieve a detection at $5\sigma$  level after trials, at least $5.8\sigma$ pre-trials excess is needed in one of the investigated locations within the location uncertainty of the event.
This example effectively is corresponding to a $\sim15\%$ worsening of the sensitivity. 
Additionally, possible drop of acceptance at higher offsets from the IACT camera centre can further worsen the performance.  

The situation is even more complicated if no significant emission is detected and one is interested in putting constraints on the emission of the event with uncertain location.
The upper limit of the emission exploits the observed signal \citep{2005NIMPA.551..493R}, namely a given flux value can be consistent with a small positive excess, but would be unlikely if null or negative excess is observed.
In the sky region covering the localisation error of the event, even if no true emission is present,  it is common to observe excesses with a statistical significance of 2--3$\sigma$, just by the combination of the Gaussian distribution of the used test statistic and the number of trials. 
The limit of the emission in those particular sky positions will be then larger, even by a factor of a few, than in the average location on the sky. 
As one of those positions might in fact hide a real, weak source, the most conservative approach for putting a limit on the TeV emission associated to the poorly localised event would be the least constraining one, which severely worsens the performance for constraining emission models.   

WCD and SA as wide FOV instruments, are more suitable for a follow-up of poorly localised alerts as they can simultaneously scan a large area in the sky.
While they are also affected by the number of trials problem in the search of such an emission, as their angular resolution is usually worse than of IACTs, the corresponding effect of trials is lesser. 
However, for the time scale of transient sources the sensitivity of WCD/SA are usually significantly worse than for IACTs (see e.g. \citealp{2017A&A...607A.115I}).  

\section{Conclusions}

TeV astronomy is undergoing its golden age. 
Most of the $\sim250$ sources currently known in this energy range had their TeV emission discovered by the current generation of instruments.  
Moreover, they cover a range of various types of objects, showing the versatility of the IACT, WCD and SA techniques. 
It is to be expected that those sources are only a tip of the iceberg of the sources available to the upcoming generation of instruments. 
While the standard analysis techniques for both IACT and WCD/SA instruments are well established, there is still an ongoing effort at improving them. 
Recently this effort concentrates on one hand on usage of the DL techniques. 
On the other hand, there is also a growing interest in development of open tools and common data formats that facilitate the analysis and allow simple combination of data sets from various experiments. 

While IACT, WCD and SA instruments operate in a similar energy range, the differences of the used techniques cause differences in their operation mode and performance parameters, that allow them to complement each other.
All those types of techniques are also desirable in the join MWL or MM studies with other instruments.  


\vspace{6pt} 



\funding{
This research was funded by Narodowe Centrum Nauki grant number 2019/34/E/ST9/00224.
}



\acknowledgments{
The author would like to thank Takashi Sako, Rub\'en L\'opez-Coto, Harm Schoorlemmer as well as anonymous journal reviewers for providing useful comments to the manuscript.
}

\conflictsofinterest{The author declares no conflict of interest.} 



\abbreviations{Abbreviations}{
The following abbreviations are used in this manuscript:\\

\noindent 
\begin{tabular}{@{}ll}
CNN & convolutional neural networks \\
CTA & Cherenkov Telescope Array \\
DL & deep learning \\
FOV & field of view \\
HAWC & High Altitude Water Cherenkov \\
H.E.S.S. & High Energy Stereoscopic System \\
IACT & Imaging Atmospheric Cherenkov Telescope \\
LHAASO & Large High Altitude Air Shower Observatory \\
LHAASO-WCDA & Water Cherenkov Detector Array \\
LHAASO-WFCTA & Wide Field air Cherenkov Telescopes array \\
LST & Large-Sized Telescope \\
MAGIC & Major Atmospheric Gamma-ray Imaging Cherenkov \\
MC & Monte Carlo (simulations) \\
MM & multi-messenger \\
MWL & multi-wavelength \\
PMT & Photomultiplier Tube \\
PSF & Point Spread Function \\
RNN & recurrent neural networks \\
SA & Surface Array \\
SiPM & Silicon Photomultiplier \\ 
SWGO & Southern Wide-field Gamma-ray Observatory \\
TOO & Target of Opportunity \\
VERITAS & Very Energetic Radiation Imaging Telescope Array System \\
VHE & Very High Energy \\
WCD & Water Cherenkov Detector 
\end{tabular}}

\appendixtitles{no} 
\appendixstart
\appendix

\collend 
\reftitle{References}


\externalbibliography{yes}
\bibliography{tev_instr_review}

\end{document}